\documentclass[aps,pra,preprint,showpacs,superscriptaddress,amsfonts,floatfix]{revtex4}
\usepackage{graphics}

\usepackage[dvips]{graphicx}
\usepackage{amsmath}

\begin{document}

%
%

\title{$Z_1$-oscillation in the retarding force of metals for slow ions: \\
Comparative study of theoretical modelings}

%
%

%
\author{I. Nagy}
\affiliation{Department of Theoretical Physics,
Institute of Physics, \\
Budapest University of Technology and Economics, \\ H-1521 Budapest, Hungary}
\affiliation{Donostia International Physics Center, P. Manuel de
Lardizabal 4, \\ E-20018 San Sebasti\'an, Spain}
\author{ I. Aldazabal}
\affiliation{Centro de F\'{i}sica de Materiales (CSIC-UPV/EHU)-MPC,
P. Manuel de Lardizabal 5, \\ E-20018 San Sebasti\'an, Spain}
\affiliation{Donostia International Physics Center, P. Manuel de
Lardizabal 4, \\ E-20018 San Sebasti\'an, Spain}

\date{\today}
\begin{abstract}

Theoretical calculations, applying elements of the orbital-based density functional method for self-consistent screening in an electron gas to the stopping power of metals for slow projectiles are reviewed, by focusing on differences in modelings.
New, two-channel-based results on the average retarding force are presented.
All theoretical results are compared with pioneering experimental 
data, obtained for carbon target, on $Z_1$-oscillation in stopping at random collisional condition.

\end{abstract}


\maketitle


\section{Introduction and Motivation}

According to the profound classification of Hawking \cite{Hawking10}, a modeling of reality in theoretical physics is good if it contains few adjustable elements, agrees with and explain existing observations, and makes detailed predictions about  future observations.  For instance, within our personal selection, in the famous treatment of Landau on extended, interacting Fermi systems the determination of the low-energy excitation spectrum requires  the introduction of just one adjustable element \cite{Kadanoff62,Migdal77}, the quasiparticle effective mass ($m^{*}$).

Hawking adopted a view that we can call model-dependent realism, which provides a framework,
generally of mathematical nature, with which to interpret physical phenomena. But 
different theories can successfully describe the same phenomenon \cite{Hawking10}. 
Moreover, many theories that  proven successful were later replaced by other ones based on new {\it concepts}. 

Motivated by such a profound view on model-dependent realism \cite{Hawking10}, here we turn to a particular phenomenon well-known in different subfield of nature. It is the energy loss of charged projectiles \cite{Landau58} in matter, and the characterization of  the slowing down by an
average retarding force experienced by heavy intruders. Evidently, the phenomenon prescribes a quantum (wave) mechanical consideration of electrons in the target material.  Furthermore, a retarding-force interpretation of the energy loss per unit path length ($dE/dx$) for ions requires the calculation of expectation values of the force operator (gradient of an external potential) considering complete sets for electron states in order to quantify an observable.

Pioneering experiments \cite{Ormrod63,Ormrod64} with slow ($v=0.41$) ions with $Z_1\in[1,20]$
on solid carbon target ($Z_2=6$, with four valence electrons) at {\it random}
collisional condition revealed a pronounced $Z_1$-oscillation
in the observable stopping power (we use Hartree atomic units throughout this work).
For an account on $Z_1$-oscillation in gases we refer to \cite{Teplova10}. 

Our present, quantum mechanical, understanding of $Z_1$-oscillations in metallic targets with slow charged projectiles dates back to \cite{Puska83,Echenique86,Nagy89}, as we will outline in the next Section. New elements to these single-channel theoretical results 
(obtained for a Kohn-Sham-type degenerate electron gas model) were already considered and quantified in \cite{Arista94,Pitarke05} and, quite recently, in \cite{Nagy20} via two (phase-shift-based)
nonlinear channels to the average retarding force. 
For a discussion of statistical, quasiclassical theoretical attempts \cite{Firsov59,Lindhard61} on the electronic stopping power
we refer to a summary \cite{Sigmund08}. Our work here is dedicated
to a comparative study on different modelings \cite{Puska83,Echenique86,Nagy89,Arista94,Pitarke05,Nagy20} of the observed $Z_1$-oscillations in carbon.


\section{Comparison and discussion}

We follow here a time-ordered path in order to  outline the above-selected 
\cite{Puska83,Echenique86,Nagy89,Arista94,Pitarke05,Nagy20} modelings
on low-velocity ($v$) stopping power. In order to concentrate on concepts, and associated new adjustable elements, we will avoid extensive (repeated) mathematical details (where it is possible without distorting the reasoning) by referring for them to the original papers. 

In a scattering interpretation the energy loss per unit path length is a simple product 
\begin{equation}
\frac{dE}{dx}\, =  n_0\, [(vv_F)\, \sigma_{tr}(v_F)]
\end{equation}
where $\sigma_{tr}$ is the transport (momentum transfer) integrated cross section\textbf{}. The electron number density (a  part of constituents of metals) is $n_0$ and the velocity in the last occupied
one-electron momentum state (plane wave) is $v_F$. In this semiclassical interpretation $\sigma_{tr}$
can be based on a probability (instead of a probability amplitude) and is defined by
\begin{equation}
\sigma_{tr}\, =\, 2\pi\, \int_{0}^{\pi}\, d\theta\, \sin\theta\, (1-\cos\theta)\, d\sigma(\theta) 
\end{equation}
in terms of the differential cross section $d\sigma(\theta)$. Quantum statistics for 
fermions is encoded in $v_F\equiv{(3\pi^3n_0)^{1/3}}$. The product $vv_F$ is a simple arithmetic average of energy change ($\omega$) in two-particle (binary) collision, i.e., 
$<\omega>=v[(2v_F+0)/2]=vv_F$.

The wave character of system electrons appears in the quantum mechanical description of $\sigma_{tr}$
by using a complex scattering amplitude $[f(\theta,v_F)]$ to $d\sigma\propto{|f(\theta,v_F)|^2}$. Thus, in terms of Bessel phase shifts in partial waves for regularized (screened) interaction one arrives at
\begin{equation}
\sigma_{tr}(v_F)=\frac{4\pi}{v_F^2}\, \sum_{l=0}^{\infty}\, (l+1)\, \sin^2[\delta_l(v_F)-\delta_{l+1}(v_F)].
\end{equation}
Pioneering papers \cite{Puska83,Echenique86} are based on the above-outlined kinetic framework.
Their remarkable contribution (an adjustable element)  was to apply phase shift outputs of the self-consistent procedure of auxiliary-orbital-based (Kohn-Sham) DFT for screening of an embedded bare charge $Z_1$, independently of incoming-ion charge states. In such a buildup-procedure of complete screening, where we respect the impact of a metallic medium in {\it all} one-electron states 
{\it a priori}, one has the Friedel sum rule of neutrality with charge $Q$ 
on \cite{Mahan81} impurity
\begin{equation}
Q\equiv{Z_1}\, =\, \frac{2}{\pi}\, \sum_{l=0}^{\infty}\, (2l+1)\,  \delta_l(v_F),
\end{equation}
and the Levinson theorem of scattering theory with {\it local} \cite{Burke77} interaction is satisfied.

Notice, that the possible role of relative (r) kinematics within an independent-particle-scattering picture beyond the impurity limit ($v\rightarrow{0}$) was investigated in thesis-papers
\cite{Remi06,Remi07} for $v\in{(0,v_F)}$, by considering prefixed (at $v\rightarrow{0}$) 
 DFT potentials \cite{Puska83,Echenique86,Nagy89} as plausible ones.
 Notable reduction in ratios, obtained at $v\rightarrow{0}$, of maximum/minimum in $Z_1$-oscillation
 was observed for $v\rightarrow{v_F}$. As expected on physical grounds, and due to mathematics in
 $\sigma_{tr}(v_r)=(4\pi/v_r^2)\sum_{l=0}^{\infty}(l+1) \sin^2[\delta_l(v_r)-\delta_{l+1}(v_r)]$,
 boundary values in oscillations are the most sensitive to tuning the relative (r) wave number.
 The comparison with surface-experimental data \cite{Winter03}, which are almost free from complications due to lattice-ions, gave an improved character for $v=0.5$ at
 Al surface ($v_F\simeq{0.87}$) in $Z_1$-oscillation.

Of course, as we know from an insightful book \cite{Peierls79}, the above-discussed 
(mean-field) consideration of screening is one possibility. An alternative way uses the
consideration that the matrix elements
of the external potential which connect occupied states described by Slater determinant has no effect
on the many-body wave function since their effect cancels out in a determinant wave function.
This  fact suggests starting from the Schr\"odinger equation in which the potential is replaced by a
truncated potential which is defined like the Bethe-Goldstone potential, except that it is concerned
with one particle only, and therefore simpler. However, such a pseudopotential is {\it nonlocal} since
in its definition for a given value of the space coordinate the value of one-particle wave function at
other space-points enter. 

We are not aware of such an implementation for screening to stopping calculations via phase shifts, but we should keep in mind this alternative way for cases of partially-ionized projectiles since there are similar values in $\sigma_{tr}$ for small phase shift differences or differences close to $\pi$. With
nonlocal \cite{Burke77} interaction the modified Levinson theorem becomes
\begin{equation}
[\delta_l(0)-\delta_l(\infty)]\, =\, \pi(n_b+n_p)  
\end{equation}
where $n_b$ is the number of true bound states and $n_p$ is the number of "bound states" 
excluded by the Pauli exclusion principle for fermions. For instance, in electron-helium atom scattering
in vacuum $n_b=0$ and $n_p=1$, thus $\delta_0\rightarrow{\pi}$ since there is no stable negative ion. In orbital-based construction of DFT with $Z_1=2$ one has $n_b=1$ and $n_p=0$ thus $\delta_0\rightarrow{\pi}$ for vanishing scattering wave number, i.e., at very low 
density \cite{Nagy94} of the electron gas. However, a similar tendency to $\pi$ does not imply, 
to our best knowledge, the same finite values for the scattering length and thus for cross sections related to observables.

Results based on \cite{Puska83,Echenique86} for $(1/v)(dE/dx)$, obtained within an orbital-based (Kohn-Sham) treatment for screening and scattering of an embedded $Z_1$ are plotted in Fig. 1 by dashed curve for 
$r_s\simeq{1.6}$ of the Wigner-Seitz parameter, $r_s=[3/(4\pi n_0)]^{1/3}=(9\pi/4)^{1/3}/v_F$.
This parameter represents the valence electron density of carbon target . It was pointed out in 
the pioneering \cite{Echenique86} that  the data (solid circles) and theory (dashed curve) show the same overall trend with good quantitative agreement for $Z_1<7$. However, as was emphasized in the
other pioneering work
\cite{Puska83}, experimental values tend to increase more rapidly as $Z_1$ increases.  According
to the qualitative discussion of physics in \cite{Puska83}, in a kinetic picture, the increase can be due to several effects. For instance, the ionic radius increases
with $Z_1$ and therefore heavier ions see a larger effective electron density than the lighter ones.

The closely related problem of charge inhomogeneity under random collisional condition in carbon, was addressed and quantified in \cite{Arista94}. There, as a necessary  extension of \cite{Lindhard61} 
with hidden details \cite{Sigmund08}, an averaging procedure was defined based on a statistical local-density approximation, with $[dE/dx](r)$ local inputs constrained by Eq.(4), via $r_s(r)$ as follows
\begin{equation}
\langle{\frac{dE}{dx}}\rangle \, \equiv{N_a\, \int_{0}^{r_a}\, \frac{dE}{dx}(r)\, 4\pi r^2 dr}.
\end{equation}
Here $N_a$ is the number of atoms per unit volume and $r_a$ is the atomic cell radius. 
The dash-dotted curve in Fig. 1 shows the such-obtained averaged results. There is a remarkable
improvement especially at around the experimental minimum and beyond it for higher $Z_1$.

However, in a more recent analysis \cite{Arista10} on averaging procedures it was pointed out
that one of the shortcomings of the above approach is the fact that it includes in the integration
all the atomic electrons ($r_{min}=0$) without consideration of the important binding effects that
will tend to cancel the contribution of inner shell(s) to the stopping in the case of slow ions. 
In the case of carbon with four valence electrons this shell is the doubly occupied $1s$ shell.
Clearly by excluding this (compact in its extension) shell from averaging one goes back to the dashed curve.
In our opinion, this alarming observation on the realistic role of target-atom inner shells remains valid for metallic targets with heavier  \cite{Arista19} atomic constituents as well. Intuitively, swift projectiles 
\cite{Nagy02} without bound electrons around them, say $Z_1=\pm{1}$, could be 
better candidates to a statistical averaging procedure as the one in Eq.(6).

\begin{figure} 
\scalebox{0.4}[0.4] {\includegraphics{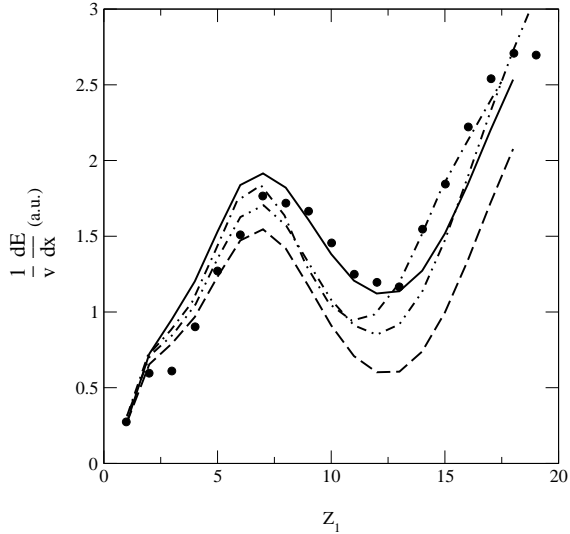}}
\caption{Stopping power of carbon ($Z_2=6$) target
($r_s\simeq{1.6}$) for ions with atomic number $Z_1$. Transmission
data (denoted by solid circles), obtained at {\it random} collisional condition, are taken from \cite{Ormrod63,Ormrod64} for $v=0.41$. 
The dashed curve corresponds to the conventional (single-channel) result obtained
in DFT at embedding condition \cite{Puska83,Echenique86}.
The dash-dotted, and dash-dot-dot-dashed curves (with new elements beyond this conventional
description) refer to \cite{Arista94} and \cite{Pitarke05}, respectively. The solid curve is based on 
Eqs.(7-9), and refers to our two-channel approach developed recently \cite{Nagy20}.           } 
\label{figure1}. 
\end{figure}
%


To DFT calculations \cite{Puska83,Echenique86} the electrons (without impurity) are considered as noninteracting in their plane-wave Kohn-Sham one-electron states.  The momentum distribution 
function of them is an ideal Fermi-Dirac distribution with (invariant) Fermi surface defined from the density $n_0$. However, as many-body methods \cite{Kadanoff62,Mahan81,Pines61} indicate, the excitation spectrum above such a
surface may get modulations (due to dynamical exchange-correlation
\cite{Nagy85} between electrons) beyond the one prescribed by the imaginary part 
of the noninteracting ($m^{*}=m$) density-density response function
[which is $\propto{\omega}$, independently of the characteristics ($v_F$) of  the filled ground-state], 
i.e., of the Lindhard function of the time-dependent  linearized Hartree-type approximation for screening. In \cite{Nagy89} we made an attempt to combine static screening nonlinearities with many-body modulation
in the electron-hole excitation spectrum (which should depends on $m^{*}$, according to Landau's physics-based \cite{Kadanoff62,Pines61} intuitive treatment)
responsible for dissipation. Only {\it small} changes in stopping, depending slightly on $Z_1\in{[5,40]}$, were observed in \cite{Echenique89} for an electron gas with $r_s=2$.


Therefore, it was a surprise when a sophisticated approach \cite{Pitarke05} based on 
time-dependent density-functional theory (TDDFT), for the prototype model system of an embedded $Z_1$ and an interacting electron gas, within its local-density approximation gave a much {\it higher} modulating effect [a density-gradient-dependent additive term to Eq.(1)] in $Z_1$-oscillation of results plotted by dashed curve. This prediction is exhibited in Fig. 1 by a dash-dot-dot-dashed curve. The theoretical enhancement was interpreted tentatively as excitonic effect in electron-hole pair excitation and good agreement with experimental data on carbon target (see, Fig. 1) at random collisional condition was concluded.
Later the above-outlined adjustable element of modeling was reviewed \cite{Pitarke07} within the physically more consistent time-dependent current density functional theory (TDCDFT). Briefly, this
new approximation, due to its tensorial character, can satisfy important conservation laws and associated sum rules. Owing to such a physical consistency, the re-adjusted numerical results show only very {\it small} changes to pioneering DFT results, i.e.,  to the dashed curve in Fig. 1.


Viewing metals as systems of degenerate electron gases (with corresponding $r_s$ parameters)
and fixed lattice ions, a recent theoretical attempt \cite{Nagy20}, with the goal to extend 
\cite{Puska83,Echenique86,Nagy89}, considered the role of sudden charge-change cycles, as an adjustable new element to the average retarding force experienced by projectiles in their slowing down under random collisional conditions. There, the force interpretation was implemented by using
quantum mechanical calculations for matrix elements, instead of direct probability interpretation behind
a differential cross section needed in the classical definition in Eq.(2). 

We believe that, within the
wave mechanical description of one-electron states, such a matrix-element-based \cite{Nagy20} treatment with an additional sudden time-dependent process is a  physically consistent one. 
Clearly, and in agreement with an enlightening analysis \cite{Lindhard71} of an esteemed expert
in the field of stopping, the {\it major} (from stopping point of view)
change from a free Fermi gas occurs because the strong Coulomb
field around lattice nuclei allow an increase in the rate of local (sudden) processes of electron
jumping into and away from bound states on the projectile. Band-related further, presumably $l$-dependent, effects are not addressed here in our projectile-centered construction of one-particle scattering states.

Technically, we used earlier established works  
\cite{Gaspari72,Bonig89,Tang98} for screened interaction energy
in order to model a new (regularized Coulombic)  adjustable dissipative channel which
could model the sudden ionizing role of lattice ions in the many-body system of electrons at random
collisional condition in transmission. Concretely, in \cite{Nagy20} we derived
\begin{equation}
\frac{1}{v}\, \frac{dE}{dx}\, =\, \left[Q^{(1)}(v_F)\, + Q^{(2)}(v_F)\right],
\end{equation}
where the two coefficients ($q\neq{0}$) are given by the following expressions
\begin{equation}
Q^{(1)}(v_F)\, =\, \frac{4}{3\pi}\, v_F^2\, \sum_{l=0}^{\infty}(l+1)\sin^2[\delta_l(v_F)-\delta_{l+1}(v_F)],
\end{equation}
\begin{equation}
Q^{(2)}(v_F)\, =\, \frac{4}{3\pi}\,  q^2\, \sum_{l=0}^{\infty}\frac{1}{l+1}\, 
\sin^4\left[\frac{\delta_l(v_F)-\delta_{l+1}(v_F)}{2}\right].
\end{equation}
We used phase shifts values from earlier works \cite{Puska83,Echenique86,Nagy89} in this two-channel
modeling  and the results ($q=1$) obtained from Eq.(7) is plotted by solid curve in Fig. 1. The agreement
with data, especially beyond the first maximum in oscillation, is very reasonable. 
Thus, we believe, that the underlying complete set of scattering states generated in DFT via an
iterative procedure by embedding a bare charge $Z_1$ is, {\it a posteriori}, the proper relaxed one
even for singly ionized external projectiles. This was expected on physical grounds by considering the immediate screening action of the charged many-body system of metals.

In order to provide a template-like figure to future new
developments, we do not plot data (for instance for Al target as in \cite{Nagy20} for $Z_1\leq{20}$, at random collisional condition) in our second, illustrative Fig. 2. This shows, in its clinical form, our
prediction with a new adjustable element to characterize an observable. We suggest further attempts.
Especially, experiments with $Z_1\in{[20,40]}$ on Al at random collisional condition are recommended. 

\begin{figure}
\scalebox{0.4}[0.4] {\includegraphics{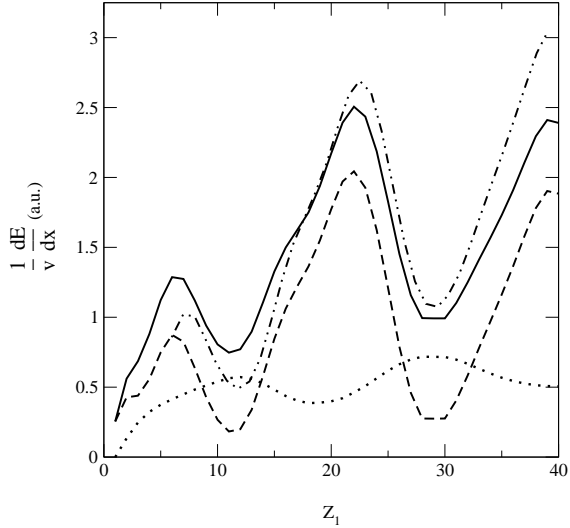}}
\caption{Illustrative theoretical stopping power at $r_s=2$ for slow ions with atomic number $Z_1$.
The dashed and dotted curves refer to Eq.(8) and Eq.(9), respectively. Their sum is plotted
by solid curve. See the text, and \cite{Nagy20}, for further details on our force-interpretation 
behind channels.} 
\label{figure2}. 
\end{figure}

We motivated our two-channel approach for an average retarding force by referring to sudden local charge changing processes behind $q=1$ at random collisional situation in metals. Can one characterize such an average within the framework of an effective single-channel modeling
via $\bar{\sigma}_{tr}(v_F)$ in Eq.(1)?
The rest of this comparative study on capabilities of adjustable elements is motivated 
by such a question. Our discussion below on this (challenging) question is 
based on quantum physics but has an intuitive character.

We return to an important message put forward in \cite{Lindhard71}. For slow moving ions
there is  stationary flow (at far distances) of electrons. It was pointed out that a total balance in that flow could {\it replace} the concept of local capture-loss processes. But no quantification of this
idea was presented. Later, in a combined paper, we quantified \cite{Nagy00} such an idea based on conservation laws, considering the case of slow He ions where charge-change may occur.
We applied, as constraint to a one-parametric model potential, 
the form derived in \cite{Zwerger97} for the dipolar backflow ($\mathcal{D}_b$) amplitude
in terms of phase shifts solely at the Fermi energy
\begin{equation}
D_b=\frac{1}{\pi}\sum_{l=0}^{\infty} (2l+1)\sin{(2\bar{\delta_l})}+
\frac{4}{\pi}\sum_{l=0}^{\infty} (l+1)^2 \sin{\bar{\delta}_l}\sin{\bar{\delta}_{l+1}}
\sin{(\bar{\delta}_l-\bar{\delta}_{l+1})}      
\end{equation}
For the very clear, pedagogically enlightening discussion of conservation laws behind the above form we refer to the original detailed research paper.
This expression reduces to the familiar Friedel sum rule of charge-screening ($D_b\Rightarrow{Z_1}$) for small phase shifts, i.e., in the first-order Born limit where $(Z_1/v_F)<<{1}$ for the Sommerfeld parameter. In perturbative cases with small charges or high densities, where we have
no bound states, the second term becomes third-order in the perturbing charge. 

\begin{table}
\caption{\label{tab:parameters}  Ratios ($\mathcal{R}_i$) of He and H stopping powers for
the metallic range of $r_s$ in a time-ordered representation. The correspondences are
$i=ENAR$ \cite{Echenique86}, $i=NES$ \cite{Nagy00}, and $i=NA$ \cite{Nagy20}.  The 3rd column refers to the amplitude of the dipolar backflow ($\mathcal{D}_b$) with DFT phase shifts based on
complete screening (Friedel sum rule) of an embedded charge $Z_1=2$. 
The 5th column shows the dipolar backflow amplitude of a forced calculation \cite{Nagy00}
at the $\mathcal{D}_b\equiv{Z_1}=2$ constraint in the longitudinal current of the slowly moving charge.
See the text for further details.}
\begin{ruledtabular} 
\begin{tabular}{ccccccccc}
$r_s$ & $\mathcal{R}_{ENAR}$ &  $\mathcal{D}_b^{ENAR}$   & $\mathcal{R}_{NES}$ 
&  $\mathcal{D}_b^{NES}$ & $\mathcal{R}_{NA}$  \\
\hline
$1.5$ & $2.44$ &   $1.12$    & $2.45$   &  $2$     & $2.56$   \\
$2.0$ & $1.59$ &   $0.80$   & $2.14$    &  $2$      & $2.23$   \\
$3.0$ & $0.81$ &   $0.27$   & $1.77$    &  $2$      & $2.37$  \\
\end{tabular}
\end{ruledtabular}
\end{table}
%


Table I contains the essence of our message, by considering the problem of $He^{+}$ and proton
[described solely by its $Q^{(1)}$] stopping ratio $\mathcal{R}$ in metals at low ion velocities.
At $r_s=1.5$, which correspond to a comparatively 
high density, the ratios are similar and all ratios are below the first-order Born value, i.e., four.
At $r_s=2$ our $\mathcal{R}_{NA}$ ratio is in accord with TDDFT calculation in \cite{Zeb13}. At low density,
 $r_s=3$, our ratio is in nice agreement with experimental prediction \cite{Markin09} for $v\leq{0.2}$ and is below the theoretical (TDDFT) value of \cite{Zeb12}, which is about $4.7$, but is essentially larger 
 than $\mathcal{R}_{ENAR}\simeq{0.8}$ obtained in DFT.
 We believe that the high ($4.7$) value is
 due to underestimation of proton stopping in TDDFT simulation. This fact was stated in \cite{Zeb12} devoted to He/H anomaly in Au. For completeness, we note that at $r_s=3$ our two-channel-based He stopping is roughly equivalent to proton stopping
calculated (in DFT) at $r_s=1.5$. We stress that this (separate) statement is also in harmony with experimental data \cite{Markin09} obtained on gold target for low ion velocity, $v\leq{0.2}$,
at which we consider only an $s$-like state to define $r_s$. In more quantitative terms, we have
$[Q^{(1)}+Q^{(2)}](r_s=3)|_{Z_1=2}\simeq{0.38}$ and 
$Q^{(1)}(r_s=1.5)|_{Z_1=1}\simeq{0.31}$ based on Table I of \cite{Nagy20}.

Quite independently of $r_s$ we have a reasonable similarity  between  $\mathcal{R}_{NES}$ 
and $\mathcal{R}_{NA}$ values, which may indicate the applicability of an adjustable element discussed in \cite{Lindhard71}, and
quantified in \cite{Nagy00}. Here we found by comparison that, in a statistical sense for random collisional situations,
we can model (reinterpret) the local capture-loss processes by using a constraint in the electronic flow around a slow heavy ion which represents a longitudinal current as well, not only a static charge.
Clearly, with moving charges in a charged many-body system one has to satisfy
a fundamental constraint prescribed by continuity equation between induced charges and the longitudinal current. Remarkably, a recent large-scale orbital-based TDDFT simulation
emphasized \cite{Correa15} that application of current-based (not only charge-based) implicit version could be a promising way for future more sophisticated extensions within the framework of first-principles methods. In the orbital-free, i.e., explicit, version of TDDFT an attempt to include current
via a dynamical kinetic-energy functional has already been presented recently \cite{Ding18,White18}.
These modelings of reality within TDDFT would fit, of course, to Hawking profound classification outlined in the Introduction.


\section{summary}

We have reviewed earlier approaches and presented new results on the average retarding force of metallic targets for slow projectiles, using pioneering experimental data on $Z_1$-oscillation in carbon at random collision condition. All theoretical approaches, selected to our comparative work, are based on different adjustable elements. These are analyzed in details in order to arrive at established conclusions. We pointed out challenging problems which need future developments in 
the important field of ion stopping in metals.


%
\begin{acknowledgments}
This work was supported partly by Project PID2019-105488GB-I00 of the Spanish Ministry of Science, Innovation, and Universities, MICINN.
\end{acknowledgments}
%


\end{document}